\documentclass[
twocolumn,
superscriptaddress,
showpacs,
amsmath,amssymb,
floatfix
]{revtex4-2}

\usepackage{xcolor}
\usepackage{multirow}
\usepackage{microtype}
\usepackage[utf8]{inputenc}
\usepackage{amsmath}
\usepackage{graphicx}
\usepackage{epstopdf}
\usepackage{dcolumn}
\usepackage{bm}
\usepackage{hyperref}
\usepackage{float}

\begin{document}

\title{Magnetism and Correlated Electrons in LaCr$_2$Ge$_2$N}

\author{Jiao-Jiao Meng}\thanks{These authors contributed equally to this work.}
\affiliation{School of Physics and Optoelectronic Engineering, Shandong University of Technology, 266 Xincun West Road, Zibo, 255000, P. R. China}
\affiliation{School of Physics, Zhejiang University, Hangzhou, 310058, P. R. China}

\author{Yu-Sen Xiao}\thanks{These authors contributed equally to this work.}
\affiliation{Key Laboratory of Magnetic Suspension Technology and Maglev Vehicle, Ministry of Education, Southwest Jiaotong University, 111 Second Ring Road, Chengdu 610031, P. R. China}
\affiliation{Tsung-Dao Lee Institute and School of Physics and Astronomy, Shanghai Jiao Tong University, Shanghai 200240, P. R. China}

\author{Gen Li}\thanks{These authors contributed equally to this work.}
\affiliation{School of Physics and Optoelectronic Engineering, Shandong University of Technology, 266 Xincun West Road, Zibo, 255000, P. R. China}
\affiliation{Beijing National Laboratory for Condensed Matter Physics and Institute of Physics, Chinese Academy of Sciences, Beijing 100190, P. R. China}

\author{Shao-Hua Liu}
\affiliation{School of Physics and Optoelectronic Engineering, Shandong University of Technology, 266 Xincun West Road, Zibo, 255000, P. R. China}

\author{Bai-Zhuo Li}
\affiliation{School of Physics and Optoelectronic Engineering, Shandong University of Technology, 266 Xincun West Road, Zibo, 255000, P. R. China}

\author{Hao Jiang}
\affiliation{School of Physics and Optoelectronics, Xiangtan University, Xiangtan 411105, P. R. China}

\author{Zhen Yu}
\affiliation{School of Physics and Optoelectronic Engineering, Shandong University of Technology, 266 Xincun West Road, Zibo, 255000, P. R. China}

\author{Yi-Qiang Lin}
\affiliation{School of Physics, Zhejiang University, Hangzhou, 310058, P. R. China}

\author{Xin-Yu Zhao}
\affiliation{School of Physics, Zhejiang University, Hangzhou, 310058, P. R. China}

\author{Qing-Chen Duan}
\affiliation{Tsung-Dao Lee Institute and School of Physics and Astronomy, Shanghai Jiao Tong University, Shanghai 200240, P. R. China}

\author{Wu-Zhang Yang}
\affiliation{School of Science, Westlake University, 18 Shilongshan Road, Hangzhou, 310064, P. R. China}

\author{Chong-Yao Zhao}
\affiliation{School of Physics and Optoelectronic Engineering, Shandong University of Technology, 266 Xincun West Road, Zibo, 255000, P. R. China}

\author{Zhi Ren}
\affiliation{School of Science, Westlake University, 18 Shilongshan Road, Hangzhou, 310064, P. R. China}

\author{Yu-Xue Mei}
\affiliation{School of Physics and Optoelectronic Engineering, Shandong University of Technology, 266 Xincun West Road, Zibo, 255000, P. R. China}

\author{Yong-Liang Chen}
\affiliation{Key Laboratory of Magnetic Suspension Technology and Maglev Vehicle, Ministry of Education, Southwest Jiaotong University, 111 Second Ring Road, Chengdu 610031, P. R. China}

\author{Rui-Dan Zhong}
\affiliation{Tsung-Dao Lee Institute and School of Physics and Astronomy, Shanghai Jiao Tong University, Shanghai 200240, P. R. China}

\author{Qing-Xin Dong}
\affiliation{Beijing National Laboratory for Condensed Matter Physics and Institute of Physics, Chinese Academy of Sciences, Beijing 100190, P. R. China}

\author{Peng-Tao Yang}
\affiliation{Beijing National Laboratory for Condensed Matter Physics and Institute of Physics, Chinese Academy of Sciences, Beijing 100190, P. R. China}

\author{Shu-Gang Tan}
\affiliation{School of Physics and Optoelectronic Engineering, Shandong University of Technology, 266 Xincun West Road, Zibo, 255000, P. R. China}

\author{Bo-Sen Wang}
\affiliation{Beijing National Laboratory for Condensed Matter Physics and Institute of Physics, Chinese Academy of Sciences, Beijing 100190, P. R. China}

\author{Huiqian Luo}
\affiliation{Beijing National Laboratory for Condensed Matter Physics and Institute of Physics, Chinese Academy of Sciences, Beijing 100190, P. R. China}

\author{Jin-Guang Cheng}
\affiliation{Beijing National Laboratory for Condensed Matter Physics and Institute of Physics, Chinese Academy of Sciences, Beijing 100190, P. R. China}

\author{Xue Ming}
\affiliation{School of Physics and Optoelectronic Engineering, Shandong University of Technology, 266 Xincun West Road, Zibo, 255000, P. R. China}

\author{Cao Wang}
\email{wangcao@sdut.edu.cn}
\affiliation{School of Physics and Optoelectronic Engineering, Shandong University of Technology, 266 Xincun West Road, Zibo, 255000, P. R. China}
\affiliation{School of Physics, Zhejiang University, Hangzhou, 310058, P. R. China}

\author{Guang-Han Cao}
\affiliation{School of Physics, Zhejiang University, Hangzhou, 310058, P. R. China}

\date{\today}

\begin{abstract}
We report the synthesis, structure and physical properties of a new quaternary nitride LaCr$_2$Ge$_2$N. The compound crystallizes in the CeCr$_2$Si$_2$C-type structure (P4/mmm), featuring distinctive Cr$_2$N square sheets within Cr$_2$Ge$_2$N block layers. Physical characterizations reveal enhanced electron correlations evidenced by a Sommerfeld coefficient substantially larger than band calculations and pressure-induced deviation from Fermi-liquid behavior. Magnetic measurements show short-range antiferromagnetic correlations developing around 460 K, followed by long-range magnetic ordering at 14 K. Additionally, subtle anomalies at 378 K suggest possible electronic ordering. First-principles calculations reveal nearly-flat Cr-3d bands near the Fermi level and predict a striped antiferromagnetic ground state. This work demonstrates how electron count variation in the CeCr$_2$Si$_2$C-type structure family leads to magnetic ordering in LaCr$_2$Ge$_2$N, contrasting with the paramagnetic behavior of LnCr$_2$Si$_2$C compounds.
\end{abstract}


\maketitle

\section{Introduction}
The discoveries of copper- and iron-based superconductors have fundamentally transformed superconductivity research \cite{Hosono-LaFeAsO-F,La-Ba-Cu-O}. These material families are both layered systems, featuring conducting layers separated by charge reservoir layers along the \emph{c}-axis\cite{ZrCuSiAs-type-structure,cuprates-structure}. Importantly, their superconducting phases emerge in proximity to antiferromagnetic states, suggesting antiferromagnetic fluctuations play a crucial role in the pairing mechanism\cite{AFMandSC}. This structural and physical insight has inspired the search for new unconventional superconductors among transition metal compounds with novel structural motifs and magnetic correlations. Among these, chromium-based materials have emerged as promising platforms for studying correlated electron physics, displaying diverse ground states ranging from superconductivity in quasi-one-dimensional K$_2$Cr$_3$As$_3$ to density-wave ordering and pressure-induced superconductivity in kagome lattice CsCr$_3$Sb$_5$\cite{K2Cr3As3,CsCr3Sb5-1}. The versatility of Cr-based compounds across different structural dimensionalities thus motivates the exploration of novel Cr-containing structures with potential for interesting physical properties.

Recently, compounds with tetragonal \emph{M}$_2$\emph{X}$_2$O layers (\emph{M} = Ti, V, Mn, Fe, Co; \emph{X} = S, Se, Te, As) have drawn considerable attention due to their diverse charge and spin ordered states. Materials containing Ti$_2$\emph{X}$_2$O layers manifest both charge-density-wave behavior and superconductivity\cite{Ti2O-review}. In the Mn-containing compounds\cite{Sr2F2Mn2Se2O,La2O3Mn2Se2}, Fe-based systems\cite{Na2Fe2Se2O,Fe2La2O3E,A2F2Fe2OQ2,Na2Fe2S2O}, and Co analogs\cite{La2Co2Se2O3}, insulating antiferromagnetic ground states are observed. V-based materials likewise exhibit rich electronic and magnetic phenomena, ranging from semiconducting behavior in CsV$_2$Se$_2$O\cite{CsV2Se2O} to metallic conduction with complex electronic ordering in Rb$_{1-\delta}$V$_2$Te$_2$O\cite{RbV2Te2O} and its derivatives\cite{V2Te2O,ZhangCrystal2024}. These \emph{M}$_2$\emph{X}$_2$O layers are structurally analogous to the Cr$_2$Si$_2$C layers in \textit{Ln}Cr$_2$Si$_2$C (\textit{Ln}= Lanthanides), the latter of which displays only Pauli paramagnetism\cite{CeCr2Si2C, LnCr2Si2C-PRB}. Notably, replacing \textit{Ln} with Th or U induces antiferromagnetic order in Cr lattice\cite{ThCr2Si2C,UCr2Si2C}, while substituting Cr with Mo leads to superconductivity at 2.2 K in ThMo$_2$Si$_2$C\cite{ThMo2Si2C}. These findings demonstrate that the ground state of Cr$_2$Si$_2$C layers is sensitive to electron filling and correlations, motivating the exploration of new members in this structural family.

Here, we report LaCr$_2$Ge$_2$N, a new quaternary compound crystallizing in the \textit{Ln}Cr$_2$Si$_2$C-type structure with distinctive Cr$_2$N square sheets. Through comprehensive characterization, we demonstrate that LaCr$_2$Ge$_2$N exhibits enhanced electron correlations evidenced by both a Sommerfeld coefficient substantially larger than the band-calculation value and pressure-induced deviation from Fermi-liquid behavior in resistivity. Magnetic measurements reveal short-range antiferromagnetic correlations developing around 460 K, followed by long-range magnetic ordering at 14 K. This magnetic behavior demonstrates that, like ThCr$_2$Si$_2$C and UCr$_2$Si$_2$C, the Cr atoms in LaCr$_2$Ge$_2$N develop magnetic moments, contrasting with the Pauli paramagnetic \textit{Ln}Cr$_2$Si$_2$C series and highlighting the critical role of electron count in determining magnetic ground states. Our findings establish LaCr$_2$Ge$_2$N as a platform for studying frustrated magnetism and electron correlations in layered materials.

\section{Experimental section}

Polycrystalline LaCr$_2$Ge$_2$N was synthesized via arc-melting method under a 1:1 N$_2$/Ar atmosphere. Starting materials included La, Cr (Alfa 99.95\%), Ge (Alfa 99.999\%) powders and N$_2$ gas (99.999\%). The stoichiometric mixture of La, Cr, and Ge powders was homogenized and cold-pressed in an Ar-filled glove box. After five melting-flipping cycles to ensure homogeneity, the melt button was sealed in an evacuated quartz tube with an alumina crucible and annealed at 1000 $^\circ\mathrm{C}$ for two weeks. Attempts to synthesize the target phase through solid-state reaction of LaN, Cr, and Ge powders at 1200 $^\circ\mathrm{C}$ were unsuccessful.

Structural characterization was performed using powder X-ray diffraction (XRD) at room temperature on a PANalytical EMPYREAN diffractometer in Bragg-Brentano geometry. The instrument was equipped with an incident-beam Ge monochromator to select Cu K$\alpha_1$ radiation ($\lambda$ = 1.54056 \AA) and a PIXcel detector. Data were collected over the range $10^\circ\leq2\theta\leq 120^\circ$ in step-scan mode with a step size of 0.02$^\circ$. The XRD pattern was refined using GSAS-II\cite{GSAS-1,GSAS-2}. Initial free refinement of all atomic parameters yielded occupancies within one standard deviation of unity (0.99-1.02). In the final refinement model, all atomic occupancies were fixed at 1.0, resulting in physically reasonable thermal parameters. Detailed crystallographic data have been deposited at the Cambridge Crystallographic Data Centre under deposition number CCDC 2492613.

Electron diffraction measurements were conducted at room temperature using a FEI Tecnai G2 F20 S-Twin transmission electron microscope to examine possible superstructure reflections. Elemental analysis was performed using a LECO ONH836 analyzer for nitrogen content determination and a LECO CS844 analyzer to probe possible carbon contamination. Physical property measurements were conducted using Quantum Design instruments: magnetic susceptibility was measured using a MPMS-3 system equipped with a high-temperature option, while resistivity and specific heat were measured using a PPMS-9 Evercool system with standard four-probe method for resistivity measurements. High-pressure transport measurements were performed separately in the palm cubic-anvil cell (CAC) apparatus with high hydrostatic pressures\cite{Pressure}.

First-principles calculations were performed using the Vienna Ab initio Simulation Package (VASP) within the generalized gradient approximation (GGA)\cite{VASP,GGA}. The experimental crystal structure was used as the initial structure, and both lattice parameters and atomic positions were fully relaxed until forces converged to less than 0.0001 eV/\AA. A plane-wave basis energy cutoff of 500 eV was employed alongside a 20 $\times$ 20 $\times$ 14 $\Gamma$-centered K-mesh for density of states calculations. The value of the Hubbard $U$ was fixed at 0 eV for Cr 3$d$ and 11.0 eV for La 4$f$ orbits\cite{La-Cr-U}. Different magnetic configurations were calculated to determine the magnetic ground state.

\begin{figure}
	\centering
	\includegraphics[width=8.5cm]{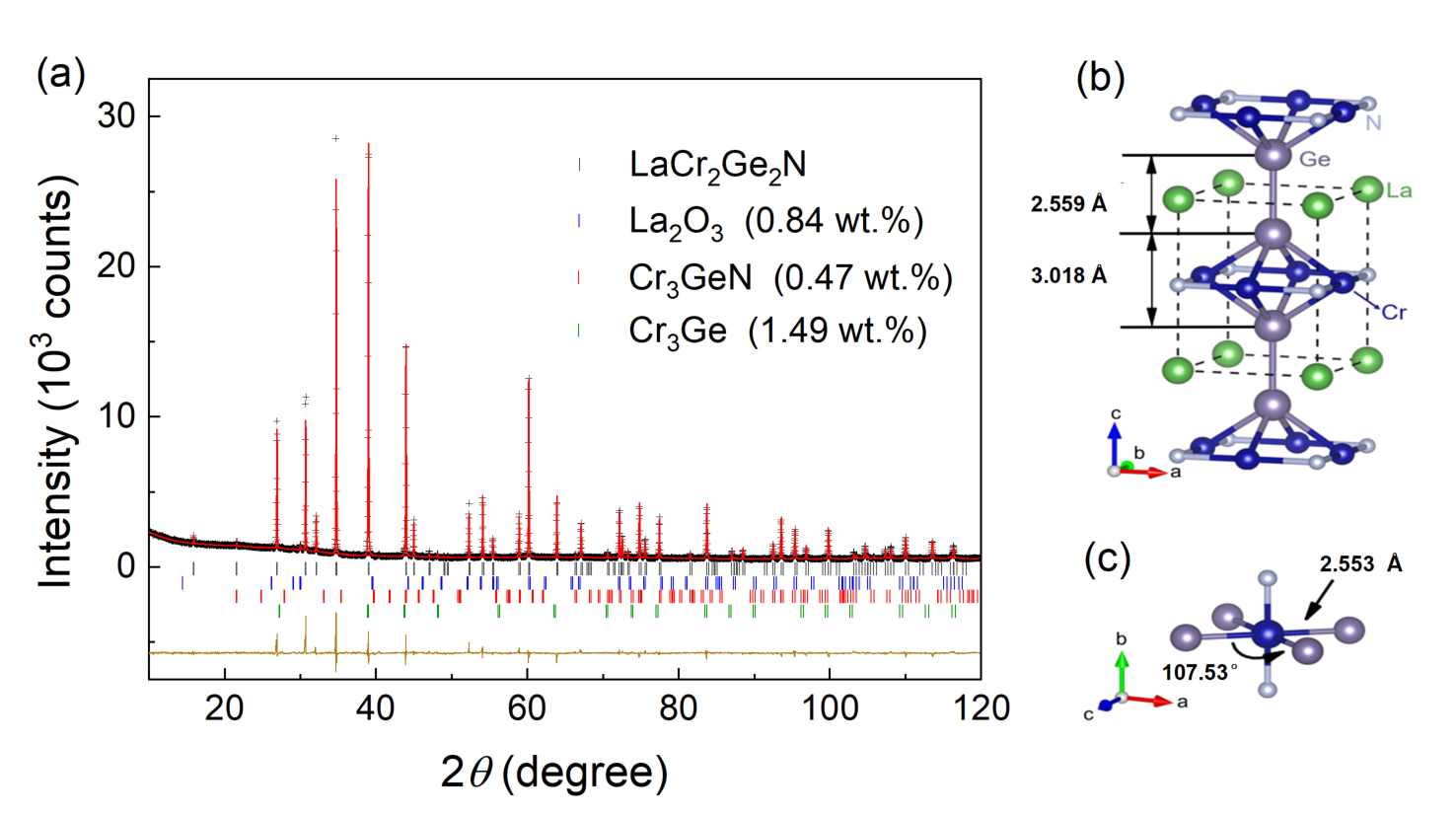}
	\caption{(a) The X-ray powder diffraction pattern of LaCr$_2$Ge$_2$N polycrystalline sample. (b) Crystal structure of LaCr$_2$Ge$_2$N. (c) Schematic of the CrGe$_4$N$_2$ octahedron.}
	\label{fig1}
\end{figure}

\begin{table}
	\caption{Crystallographic data of LaCr$_2$Ge$_2$N at room temperature.}
	\label{tab1}
	\renewcommand\arraystretch{1.2}
	\begin{center}
		\begin{tabular*}{\columnwidth}{@{\extracolsep{\fill}}ccccc}
			\hline
\hline
			&chemical formula && LaCr$_2$Ge$_2$N& \\
			&space group && $P4/mmm$& \\
			&$a$ (\AA) & &4.1183(1)& \\
			&$c$ (\AA) & &5.5768(3) &\\
			&$V$ (\AA$^3$)& & 94.584(7) &\\
			&$d_{\text{Ge-Ge}}$ (\AA)& & 2.559(2) &\\
			&$R_{\rm{wp}}$ (\%) && 4.99& \\
                        &$\chi^2$                     && 2.63&\\	
			&$Z$ && 1 &\\
		\end{tabular*}
		\vspace{0mm}
		
		\begin{tabular*}{\columnwidth}{@{\extracolsep{\fill}}lcccccl}
			\hline
			atom & site & $x$ & $y$ & $z$ & Occ.(fixed) & $B_{\rm{iso}}$ (\AA$^2$) \\
			\hline
			La & 1a &   0     & 0    & 0                 & 1.0 & 0.37(2) \\
			Cr & 2e &   0    & 0.5 & 0.5            &1.0 & 0.64(3)\\
			Ge & 2h & 0.5 & 0.5 & 0.2294(2) & 1.0 &0.64(3)\\
			N & 1b  & 0      & 0     & 0.5            & 1.0 & 0.92(5) \\
			\hline
\hline
		\end{tabular*}
	\end{center}
\end{table}

\section{Results and discussion}

Figure 1 shows the XRD pattern and Rietveld refinement results of LaCr$_2$Ge$_2$N. The \emph{PDF-4} database analysis identified three minor impurity phases: La$_2$O$_3$ (0.84 wt.\%), Cr$_3$GeN (0.47 wt.\%), and Cr$_3$Ge (1.49 wt.\%), likely originating from oxygen contamination of the starting materials. The main phase LaCr$_2$Ge$_2$N adopts the CeCr$_2$Si$_2$C-type structure ($P4/mmm$) with lattice parameters $a= 4.1183$ \AA\  and $c=5.5768$ \AA. Elemental analysis confirmed the nitrogen content of 3.39 ± 0.04 wt.\% (average of three measurements), in excellent agreement with the theoretical value of 3.41 wt.\% calculated from the phase composition determined by XRD analysis, with only trace carbon detected (0.06 ± 0.02 wt.\%). Initial Rietveld refinement with all atomic parameters freely refined yielded occupancies within one standard deviation of unity (0.99-1.02), indicating no significant vacancies. In the final refinement model, all atomic occupancies were fixed at 1.0. Detailed structural parameters are listed in Table \uppercase\expandafter{\romannumeral1}. The crystal structure of LaCr$_2$Ge$_2$N and key octahedral geometry are illustrated in Figures \ref{fig1}(b) and (c), respectively. The structure consists of alternating Cr$_2$Ge$_2$N layers and La planes in a tetragonal lattice. The Ge-Ge distance across the La plane is 2.559 \AA, slightly larger than the Ge-Ge bond length in diamond-structured germanium (2.45 \AA)\cite{Ge-Ge}, suggesting a bond order somewhat less than unity.

The structural properties of LaCr$_2$Ge$_2$N are compared with isostructural Cr-based compounds in Table \uppercase\expandafter{\romannumeral2}. These compounds share the same crystal structure with similar Cr coordination environments: Cr-$X$ bond lengths range from 2.44 to 2.56 \AA\ and $X$-Cr-$X$ bond angles range from 101$^\circ$ to 111$^\circ$. This structural similarity provides a basis for understanding their electronic structures within a common framework. However, electron counting reveals important differences: LaCr$_2$Ge$_2$N, ThCr$_2$Si$_2$C and UCr$_2$Si$_2$C have different electron counts compared to the \textit{Ln}Cr$_2$Si$_2$C series, affecting their Cr-3$d$ band filling. Indeed, while the \textit{Ln}Cr$_2$Si$_2$C series remains paramagnetic, both ThCr$_2$Si$_2$C and UCr$_2$Si$_2$C develop antiferromagnetic order in the Cr sublattice\cite{ThCr2Si2C,UCr2Si2C}. Moreover, subtle structural variations in $X$-Cr-$X$ bond angles and $c/a$ ratios among the magnetic compounds (LaCr$_2$Ge$_2$N, ThCr$_2$Si$_2$C and UCr$_2$Si$_2$C) could modulate exchange interactions and stabilize different magnetic ground states, as our first-principles calculations will demonstrate.

\begin{table*}[htbp]
	\label{tab2}
	\caption{Comparison of structural properties of 1221-type carbides and LaCr$_2$Ge$_2$N with Cr$_2X_2Y$ layers ($X$=Si, Ge; $Y$=C, N). The fractional coordinate of $X$ is (1/2, 1/2, $z_X$). $X$-Cr-$X$ denotes the bond angle within the Cr$X_4Y_2$ octahedron, Cr-$X$ is the bond length between Cr and $X$ atoms, while $X$-$X$ represents the interlayer bond length between $X$ atoms across the \textit{Ln}/\textit{An} plane, as shown in Figure \ref{fig1}(b) and (c).}
	\setlength{\tabcolsep}{3mm}
	\renewcommand\arraystretch{1.5}
	\centering
		\begin{tabular}{ccccccccc}
			\hline
\hline
			Compounds & $a$({\AA}) & $c$(\AA) & $z_X$ & $c/a$ & $X$-Cr-$X$ (°) & $X$-$\textit{}X$ (\AA) & Cr-$X$ (\AA) & Refs. \\
			\hline
			LaCr$_2$Si$_2$C & 4.048 & 5.381 & 0.235 & 1.329 & 109.63 & 2.527 & 2.477 & \cite{LnCr2Si2C-PRB} \\
			CeCr$_2$Si$_2$C & 4.014 & 5.287 & 0.224 & 1.317 & 107.97 & 2.369 & 2.481 & \cite{CeCr2Si2C} \\
			PrCr$_2$Si$_2$C & 4.022 & 5.352 & 0.227 & 1.331 & 108.42 & 2.452 & 2.479 & \cite{PrCr2Si2C} \\
			NdCr$_2$Si$_2$C & 4.006 & 5.320 & 0.211 & 1.328 & 105.07 & 2.239 & 2.521 & \cite{LnCr2Si2C-PRB} \\
			SmCr$_2$Si$_2$C & 3.985 & 5.279 & 0.228 & 1.325 & 108.42 & 2.406 & 2.456 & \cite{LnCr2Si2C-PRB} \\
			GdCr$_2$Si$_2$C & 3.983 & 5.263 & 0.224 & 1.321 & 106.44 & 2.358 & 2.465 & \cite{LnCr2Si2C-PRB} \\
                       	TbCr$_2$Si$_2$C & 3.974 & 5.244 & 0.228 & 1.320 & 101.56 & 2.374 & 2.564 & \cite{LnCr2Si2C-PRB} \\
			 YCr$_2$Si$_2$C & 3.969 & 5.219 & 0.227 & 1.315 & 108.61 & 2.368 & 2.444 & \cite{LnCr2Si2C-PRB} \\
                        DyCr$_2$Si$_2$C & 3.964 & 5.214 & 0.223 & 1.315 & 107.93 & 2.321 & 2.450 & \cite{LnCr2Si2C-PRB} \\
			HoCr$_2$Si$_2$C & 3.958 & 5.213 & 0.224 & 1.317 & 107.96 & 2.335 & 2.447 & \cite{LnCr2Si2C-PRB} \\
			ErCr$_2$Si$_2$C & 3.951 & 5.191 & 0.224 & 1.314 & 103.97 & 2.326 & 2.440 & \cite{LnCr2Si2C-PRB} \\
			ThCr$_2$Si$_2$C & 4.061 & 5.292 & 0.236 & 1.303 & 111.02 & 2.498 & 2.465 & \cite{ThCr2Si2C} \\
			UCr$_2$Si$_2$C & 3.983 & 5.160 & 0.225 & 1.296 & 109.00 & 2.319  & 2.446 & \cite{UCr2Si2C} \\
			LaCr$_2$Ge$_2$N & 4.1183 & 5.5768 & 0.2294 & 1.354 & 107.53 & 2.559 & 2.553 & This work \\
			\hline
\hline
		\end{tabular}
	\vspace{-0.5\baselineskip}
\end{table*}

\begin{figure}[tbp]
	\centering
	\includegraphics[width=8cm]{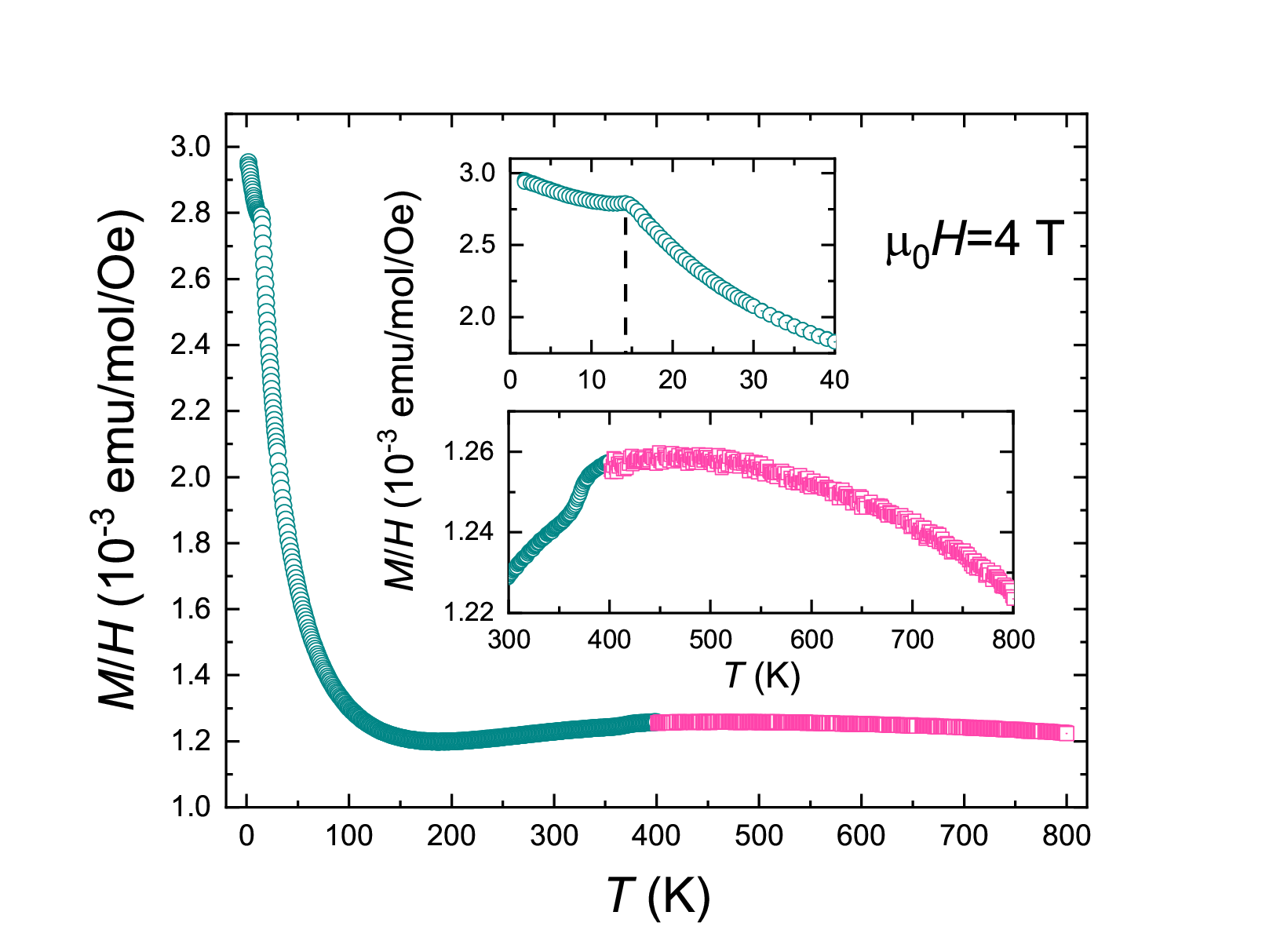}
	\caption{Temperature dependence of magnetic susceptibility $M/H$ for LaCr$_2$Ge$_2$N measured at $\mu_0H = 4$ T. The upper inset shows the low-temperature region highlighting the anomaly at $\sim$14 K. The lower inset displays the full temperature range up to 800 K, revealing a broad maximum around 460 K characteristic of short-range antiferromagnetic correlations.}
	\label{fig2}
\end{figure}

Figure \ref{fig2} presents the temperature-dependent magnetic susceptibility of LaCr$_2$Ge$_2$N measured at $\mu_0H = 4$ T. The susceptibility exhibits a broad maximum around 460 K (lower inset), characteristic of frustrated magnetic systems where competing interactions lead to the development of short-range correlations over a wide temperature range. Such behavior has been observed in various geometrically or exchange-frustrated materials, where magnetic correlations develop gradually rather than at a sharp transition.\cite{frustrated square-lattice, ThCrAsN1-xOx} The high-temperature susceptibility deviates significantly from Curie-Weiss law, indicating the presence of strong magnetic correlations rather than simple paramagnetic behavior.

At low temperatures (upper inset), the susceptibility shows an anomaly at approximately 14 K, suggesting a magnetic phase transition. Several observations support the intrinsic nature of this transition: (i) the high-field differential method employed to minimize ferromagnetic impurity contributions yields results consistent with direct measurements; (ii) the anomaly coincides with a clear feature in specific heat measurements (see below); and (iii) the known impurity phases identified by XRD (La$_2$O$_3$, Cr$_3$GeN and Cr$_3$Ge, totaling $\sim$2.8 wt.\%) do not exhibit magnetic transitions near this temperature\cite{3-impurities}. Based on these observations, we attribute this transition to the antiferromagnetic ordering of the LaCr$_2$Ge$_2$N main phase at $T_N$ = 14 K. However, we note that contributions from weakly crystalline or amorphous magnetic phases below XRD detection limits cannot be completely excluded. In addition, a careful examination of the high-temperature data reveals a subtle anomaly near 378 K, manifested as a change in the temperature dependence of susceptibility. This feature, which also appears in resistivity measurements (see below), suggests a possible electronic ordering or structural effects at high temperatures, though its exact nature remains to be determined.

\begin{figure}[tbp]
	\centering
	\includegraphics[width=8cm]{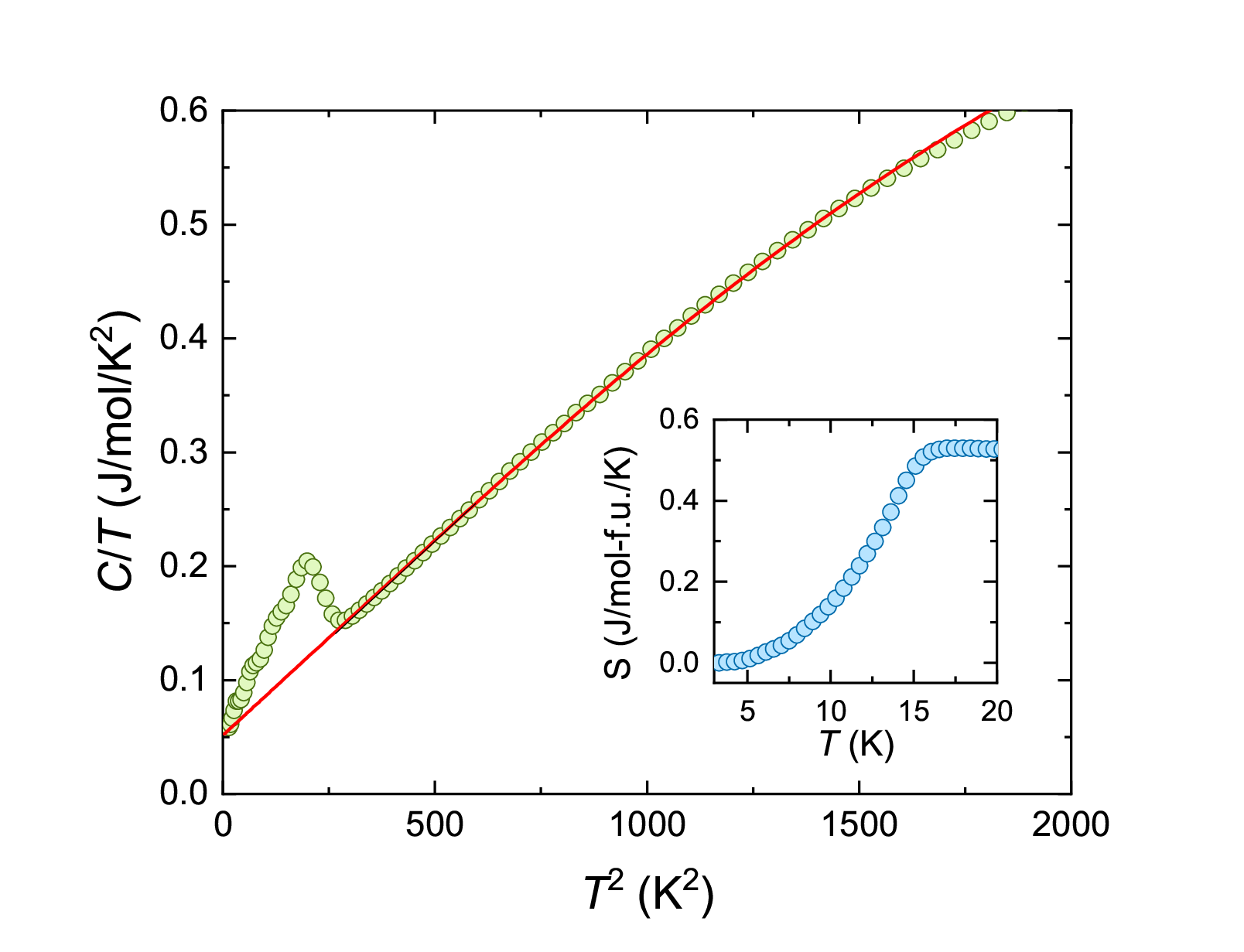}
	\caption{Low-temperature specific heat of LaCr$_2$Ge$_2$N plotted as $C/T$ versus $T^2$. The red solid line represents the fit using $C(T) = \gamma T + C_{\rm Debye}(T)$ in the range 18-40 K. The inset shows the temperature dependence of estimated magnetic entropy change obtained by integrating $C_{\rm mag}/T$.}
	\label{fig3}
\end{figure}

Figure \ref{fig3} presents the low-temperature specific heat data of LaCr$_2$Ge$_2$N, plotted as $C/T$ versus $T^2$. A clear upturn begins around $T^2 = 279$ K$^2$ (corresponding to approximately 16.7 K) and reaches a maximum near $T^2 = 196$ K$^2$ (approximately 14 K). This specific heat anomaly precisely coincides with the magnetic transition observed in susceptibility measurements, confirming its bulk nature. To analyze the electronic contribution, we fitted the data in the temperature range of 18-40 K using the model $C(T) = \gamma T + C_{\text{Debye}}(T)$. However, given that this fitting range is well above $T_N$ and may include contributions from short-range magnetic correlations, the extracted value of $\gamma$ = 51-53 mJ/mol-f.u./K$^2$ should be treated as an upper bound rather than a definitive measurement. The actual Sommerfeld coefficient could be substantially lower. Nevertheless, even with this caveat, the fitted $\gamma$ substantially exceeds the band-structure value $\gamma_{\text{band}}$ = 10.6 mJ/mol-f.u./K$^2$, indicating significant many-body effects and strong electron correlations in LaCr$_2$Ge$_2$N.

By subtracting the fitted background, we estimate the magnetic entropy change. The value of $S_{\text{mag}} \approx$ 0.53 J/mol-f.u./K obtained by integrating $C_{\text{mag}}/T$ should be considered a lower limit, as magnetic contributions from short-range correlations developing above $T_N$ may extend over a broader temperature range (up to $\sim$460 K, as suggested by the susceptibility maximum) than captured by our model. This reduced entropy—significantly smaller than $R\ln(2S+1)$ expected for fully localized magnetic moments—is consistent with substantial magnetic degrees of freedom being activated at elevated temperatures through the development of short-range correlations. The combination of enhanced electronic specific heat and reduced magnetic entropy reveals the complex interplay between itinerant and localized behaviors of the Cr 3$d$ electrons in this frustrated magnetic system.

\begin{figure}[tbp]
	\centering
	\includegraphics[width=8cm]{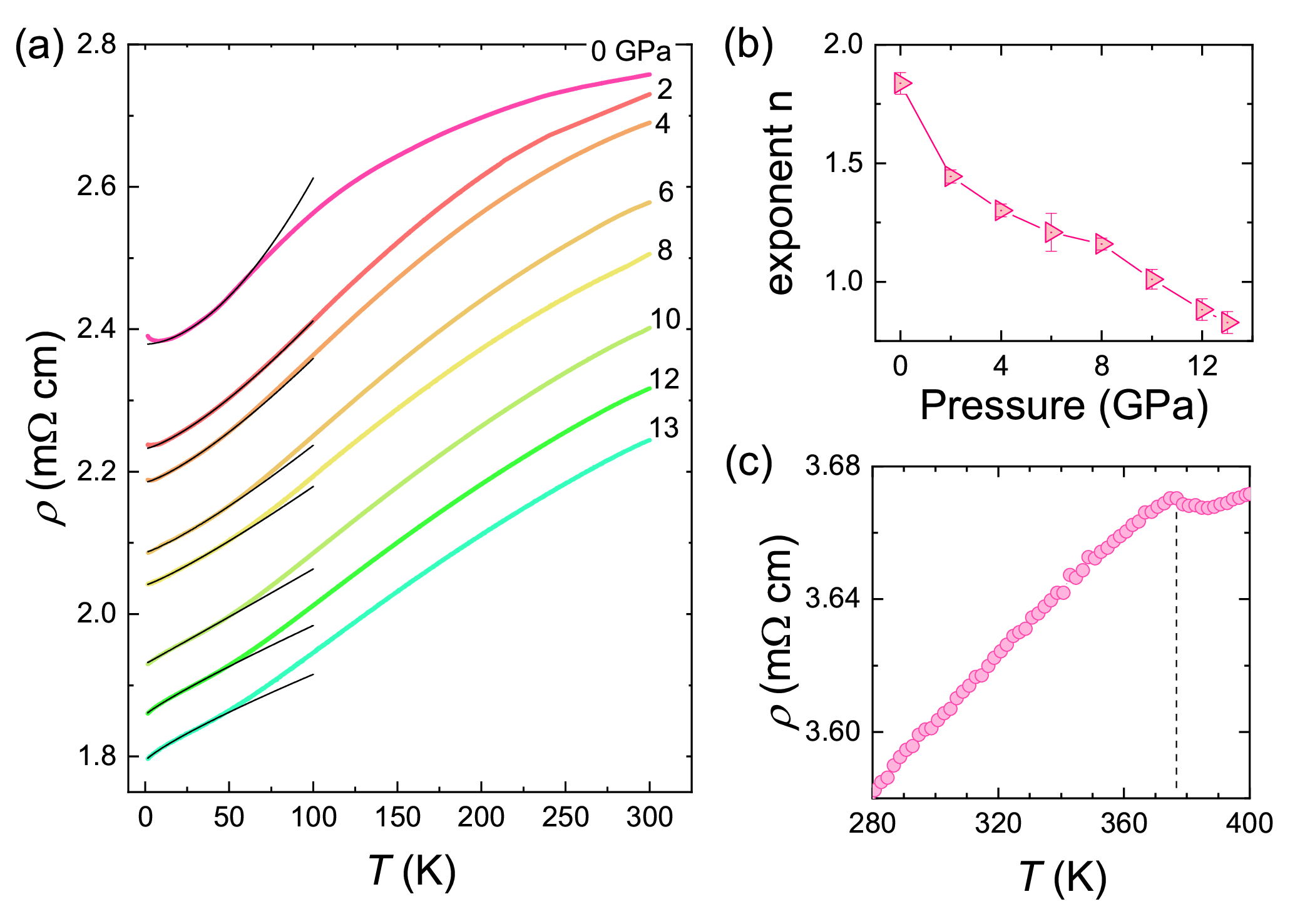}
	\caption{Temperature dependence of electrical resistivity for LaCr$_2$Ge$_2$N. (a) Resistivity measured under various pressures up to 13 GPa, with black solid lines showing power-law fits ($\rho = \rho_0 + AT^n$) at low temperatures. (b) Pressure dependence of the power-law exponent $n$. (c) High-temperature resistivity at ambient pressure, revealing an anomaly near 378 K that coincides with the susceptibility feature.}
	\label{fig4}
\end{figure}

Figure 4(a) shows the temperature and pressure dependence of resistivity for LaCr$_2$Ge$_2$N. The sample exhibits metallic behavior throughout the measured temperature range. A slight upturn in resistivity is observed below 6 K at ambient pressure and 2 GPa, which becomes suppressed at higher pressures. This low-temperature upturn is commonly observed in polycrystalline samples and can be attributed to poor intergrain contact or grain boundary effects. The suppression under pressure suggests that compression improves the electrical contact between grains. Notably, the resistivity shows no significant anomaly near 14 K at any pressure, which is consistent with the pre-existing short-range magnetic correlations (evident from the susceptibility maximum at 460 K) dominating electron scattering processes well above $T_N$.

We fitted the resistivity using $\rho = \rho_0 + AT^n$ in the temperature range of 6-45 K at ambient pressure (to avoid the upturn area) and 2-45 K under pressure, as shown by the black solid lines in Fig. 4(a). The power-law exponent $n$ decreases monotonically with increasing pressure [Fig. 4(b)], from $n\approx1.84$ at ambient pressure to $n\approx0.83$ at 13 GPa. While $n\approx1.84$ is close to the Fermi-liquid value of 2, the decrease toward $n<1$ at high pressure indicates deviation from conventional Fermi-liquid behavior. Such behavior has been observed in systems with enhanced magnetic fluctuations\cite{Non-Fermi}.

In addition, high-temperature resistivity measurements show a subtle anomaly near 378 K [Fig. 4(c)], which coincides with the feature observed in our magnetic susceptibility measurements. This correspondence suggests a possible intrinsic electronic or structural transition at this temperature. Electron diffraction measurements at room temperature did not reveal additional diffraction spots. However, precedents exist where charge-density-wave signatures were not immediately apparent in initial electron diffraction studies but were later confirmed through more sensitive techniques, as demonstrated in BaTi$_2$As$_2$O where subtle structural distortions associated with charge ordering required advanced characterization methods to detect\cite{BaTi2As2O-NC,BaTi2As2O-PRB}. Therefore, further characterization would be needed to establish the nature of the 378 K anomaly definitively.

\begin{figure}[t]
\centering
\includegraphics[width=8cm]{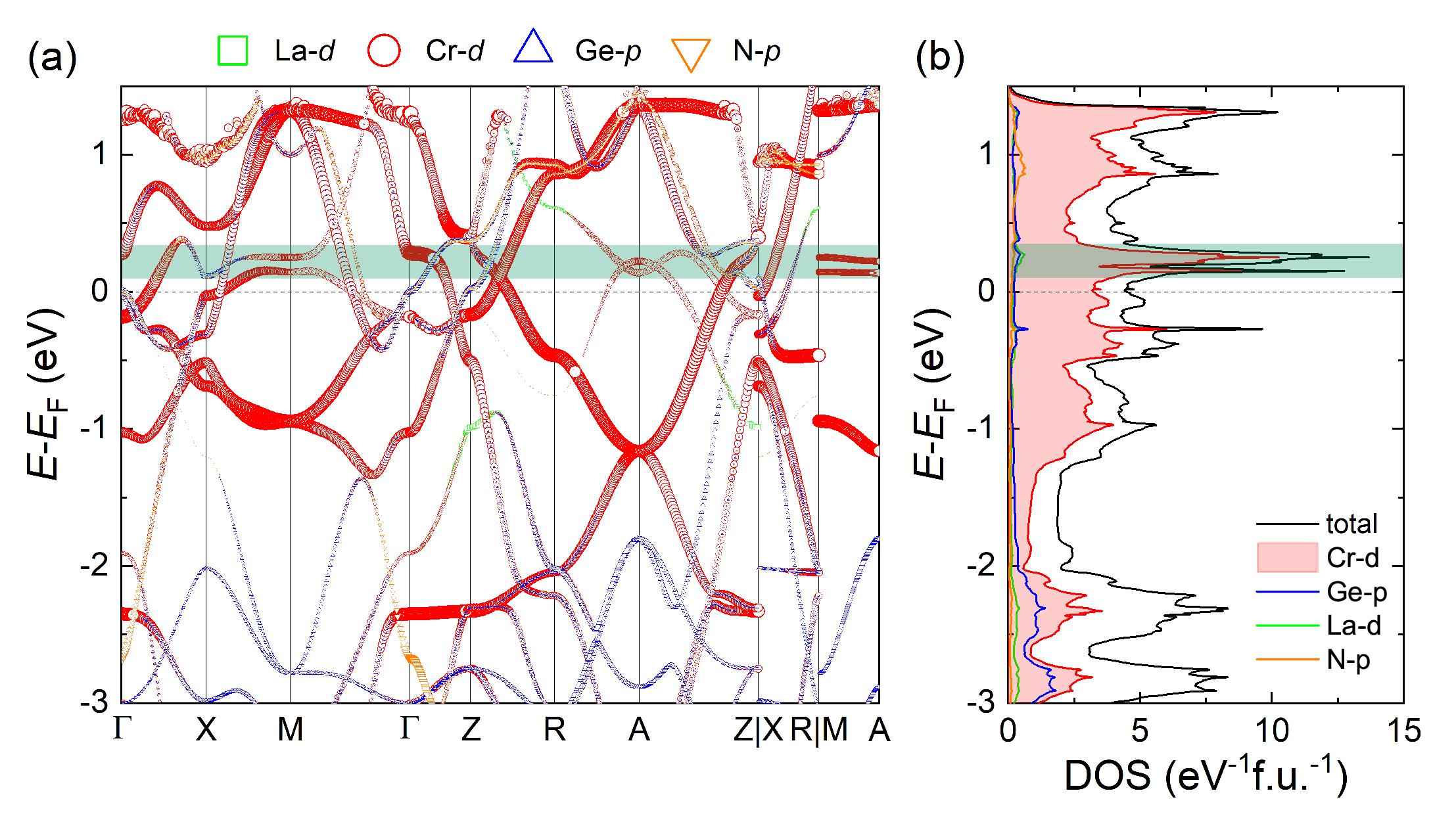}
\caption{(a) Orbital-projected band structure of LaCr$_2$Ge$_2$N along high-symmetry paths. (b) The corresponding total and orbital-projected density of states.}
\label{fig5}
\end{figure}

The reduced entropy observed in specific heat measurements, together with the enhanced Sommerfeld coefficient, suggests a complex interplay between localized and itinerant behaviors of Cr 3$d$ electrons. To elucidate this electronic structure, we performed first-principles calculations within the framework of density functional theory. Figure \ref{fig5}(a) presents the orbital-projected band structure along high-symmetry paths in the Brillouin zone. The electronic states near the Fermi level are primarily derived from Cr-3$d$ orbitals (red circles). Of particular interest are several nearly-flat Cr-$d$ bands around the M point along M-X, M-$\Gamma$, and M-A directions at about 0.15-0.25 eV above $E_F$, which give rise to a pronounced peak in the density of states [Fig. \ref{fig5}(b)]. These narrow bands could be the origin of enhanced electron correlations observed, similar to the case in Cr-based kagome system\cite{CsCr3Sb5-1,CsCr3Sb5-2}.

At the Fermi level, the calculated density of states yields $N(E_F) \approx 4.5$ states/eV/f.u., predominantly of Cr-3$d$ character. From this, we estimate the band-structure Sommerfeld coefficient $\gamma_{\rm band}$ = 10.6 mJ/mol-f.u./K$^2$. Even treating the fitted $\gamma$ according to the specific heat as an upper bound, the substantial enhancement over the band calculation indicates significant electron correlations beyond the mean-field description. These results, together with the structural similarity shown in Table \uppercase\expandafter{\romannumeral2}, allow us to understand the difference between LaCr$_2$Ge$_2$N and the \textit{Ln}Cr$_2$Si$_2$C series. Within a rigid band picture, the former has a higher electron count per formula unit, placing its Fermi level closer to these characteristic flat bands by approximately 0.2 eV [estimated from the calculated $N(E_F)$]. This proximity appears crucial for both enhancing electron correlations and stabilizing local magnetic moments on Cr sites—features that are absent in the \textit{Ln}Cr$_2$Si$_2$C series where the Fermi level lies further from these bands.

\begin{table}[tbp]
\caption{Calculated relative total energies and Cr magnetic moments for LaCr$_2$Ge$_2$N with different magnetic configurations. NM: non-magnetic; FM: ferromagnetic; A-type: antiferromagnetic between adjacent layers but ferromagnetic within layers; Striped: antiferromagnetic in one in-plane direction but ferromagnetic in the other; C-type: antiferromagnetic within layers but ferromagnetic between layers; G-type: antiferromagnetic coupling between nearest neighbors in all directions. The energies are given relative to the non-spin-polarized state ($E_{\rm NSP}$). The magnetic moments are aligned either parallel to the \emph{a}-axis or \emph{c}-axis.}

\label{tab3}
\begin{center}
\setlength{\tabcolsep}{1.1mm}
\renewcommand\arraystretch{1.2}
\begin{tabular*}{\columnwidth}{@{\extracolsep{\fill}}ccccc}
\hline
\hline
\begin{tabular}[c]{@{}c@{}}Magnetic\\ configuration\end{tabular} &Spin direction &
\begin{tabular}[c]{@{}c@{}}$E-E_{\rm{NSP}}$\\ (meV/f.u.)\end{tabular} &
$m_{\rm{Cr}}$/$\mu_B$ \\ \hline
NM     & --   & 0       & 0      \\ \hline
\multirow{2}{*}{FM} & $\|$$a$ & $-97.451$ & 1.2445 \\
       & $\|$$c$ & $-97.02$  & 1.2555 \\ \hline
\multirow{2}{*}{A-type} & $\|$$a$ & $-105.05$ & 1.198  \\
       & $\|$$c$ & $-105.06$ & 1.1985 \\ \hline
\multirow{2}{*}{Striped} & $\|$$a$ & $-172.93$ & 1.4125 \\
        & $\|$$c$ & $-172.94$ & 1.4115 \\ \hline
\multirow{2}{*}{C-type} & $\|$$a$ & $-26.337$  & 1.6795 \\
        & $\|$$c$ & $-0.0378$ & 0.0265  \\ \hline
\multirow{2}{*}{G-type} & $\|$$a$ & $-73.333$  & 0.901 \\
        & $\|$$c$ & $-3.6381$ & 0.3085  \\
\hline
\hline
\end{tabular*}
\vspace{-1.5\baselineskip}
\end{center}
\end{table}

To examine the possible magnetic ground state and correlate with our experimental observations, we calculated the total energies of various magnetic configurations. As shown in Table \ref{tab3}, all magnetic configurations considered here have lower energies than the non-magnetic state, confirming the system's tendency toward magnetic ordering, consistent with our susceptibility and specific heat measurements. Among these magnetic states, the striped antiferromagnetic configuration is energetically favored with the lowest energy of $-173$ meV/f.u. The C-type configuration shows the highest energy ($-26$ meV/f.u.) among magnetic states, highlighting the significance of in-plane magnetic arrangements in this system. The comparable energies of FM ($-97$ meV/f.u.) and A-type ($-105$ meV/f.u.) configurations suggest relatively weak interlayer coupling, which may contribute to the quasi-two-dimensional character of the magnetic correlations. The stabilization of striped magnetic order, which is also seen in the parent compounds of iron-based superconductors, can be roughly understood in terms of the $J_1$-$J_2$ model\cite{Johnston review paper}. In this context, the relatively low transition temperature observed at 14 K compared to the temperature scale of magnetic correlations (460 K) may be due to the interaction frustrations. The calculated Cr moment in the ground state is 1.4 $\mu_{\rm B}$, smaller than the expected value for localized Cr$^{3+}$ ions ($S = 3/2$), indicating substantial hybridization between Cr-3$d$ states and neighboring Ge/N atoms.

The prediction of a magnetic ground state in LaCr$_2$Ge$_2$N contrasts sharply with the Pauli paramagnetism of \textit{Ln}Cr$_2$Si$_2$C, despite similar crystal structures. This fundamental difference stems from electron count differences: LaCr$_2$Ge$_2$N has a higher electron count that positions its Fermi level closer to the nearly-flat Cr-3$d$ bands, enhancing correlations and stabilizing local Cr moments. While ThCr$_2$Si$_2$C and UCr$_2$Si$_2$C with comparable electron counts also show antiferromagnetic ordering, their magnetic structures may differ from our predicted striped configuration for LaCr$_2$Ge$_2$N. These variations highlight the delicate balance between competing exchange interactions, influenced by subtle differences in structural parameters and orbital hybridization.

\section{Conclusions}
In summary, we have synthesized a new quaternary nitride LaCr$_2$Ge$_2$N with distinctive Cr$_2$N square sheets. Physical property characterizations reveal multiple phase transitions, including a magnetic ordering transition at 14 K observed in both susceptibility and specific heat and a possible electronic transition near 378 K suggested by concurrent anomalies in both resistivity and magnetic susceptibility. The magnetic susceptibility exhibits a broad maximum around 460 K, indicating the development of short-range antiferromagnetic correlations over a wide temperature range, which is characteristic of frustrated magnetic systems. Under pressure, resistivity evolves from near-Fermi-liquid toward non-Fermi-liquid behavior, suggesting enhanced magnetic fluctuations with increasing pressure. First-principles calculations predict a striped antiferromagnetic ground state for LaCr$_2$Ge$_2$N. The specific heat analysis reveals a Sommerfeld coefficient substantially larger than band calculations, indicating significant electron correlations beyond mean-field approximations. The emergence of magnetism in LaCr$_2$Ge$_2$N, in contrast to the paramagnetic behavior of \textit{Ln}Cr$_2$Si$_2$C compounds, can be attributed to differences in electron configuration between these structurally similar systems. The simultaneous observation of enhanced electron correlations and frustrated magnetic behavior suggests a rich electronic physics in this system. These characteristics establish LaCr$_2$Ge$_2$N as a valuable platform for studying the relationship between electronic structure and magnetic frustration in layered correlated materials.

\section{acknowledgement}

This work was supported by the National Natural Science Foundation of China (Grant Nos. 12104260, 12404159), the National Key Research and Development Program of China (Grant Nos. 2023YFA1406101, 2022YFA1403202), the Natural Science Foundation of Shandong Province, China (Grant Nos. ZR2023MA028, ZR2024QA238), and Beijing National Laboratory for Condensed Matter Physics (contract no. 2023BNLCMPKF018). The authors acknowledge the support from the CAC station of Synergetic Extreme Condition User Facility (SECUF).

\end{document}